\documentclass[
reprint,
amsmath,amssymb,
aps,
pre,
]{revtex4-1}

\usepackage{amsmath}
\usepackage{cag}       
\usepackage{dcolumn}   
\usepackage{graphicx}  
\usepackage{lipsum}    
\usepackage{microtype} 
\usepackage{physics}   
\usepackage{paralist}  
\usepackage{pgfplots}  
\usepackage[caption=false]{subfig}
\usepackage{siunitx}   
\usepackage{tikz}      
\usepackage{xcolor}    
\usepackage[disable]{todonotes}   
\usepackage[breaklinks]{hyperref} 
\usepackage[capitalise]{cleveref} 

\pgfplotsset{compat=1.4}
\pgfplotsset{major grid style = {dashed, gray!40!white}}

\definecolor{accentred}{RGB}{158,11,15}

\crefname{appendix}{appendix}{appendices}
\Crefname{appendix}{appendix}{appendices}

\newcommand{\bubble}{microsphere}
\newcommand{\Bubble}{Microsphere}
\newcommand{\bubbles}{\bubble s}
\newcommand{\Bubbles}{\Bubble s}
\newcommand{\GF}{\ensuremath{G_r(\vb{x}, t; \, \vb{x}', t')}}
\DeclareRobustCommand{\redTriangle}{\textcolor{red!75!black}{\tikz{\pgfuseplotmark{triangle*}}}}
\DeclareRobustCommand{\blueCircle}{\textcolor{blue!75!black}{\tikz{\pgfuseplotmark{*}}}}

\begin{document}

\preprint{APS/123-QED}

\title{Computational dynamics of acoustically-driven \bubble\ systems}

\author{Connor Glosser}
 \email{glosser1@msu.edu}
 \homepage{http://www.msu.edu/~glosser1}
 \altaffiliation[Also at ]{Michigan State University, Department of Electrical \& Computer Engineering.}
\author{Carlo Piermarocchi}
 \email{carlo@pa.msu.edu}
\affiliation{
  Michigan State University,\\
  Department of Physics \& Astronomy
}

\noaffiliation

\author{Jie Li}
\author{Dan Dault}
\author{B.\@ Shanker}
\affiliation{
  Michigan State University,\\
  Department of Electrical \& Computer Engineering
}

\date{\today}

\begin{abstract}
We propose a computational framework for the self-consistent dynamics of a \bubble\ system driven by a pulsed acoustic field in an ideal fluid.
Our framework combines a molecular dynamics integrator describing the dynamics of the \bubble\ system with a time-dependent integral equation solver for the acoustic field that makes use of fields represented as surface expansions in spherical harmonic basis functions.
The presented approach allows us to describe the inter-particle interaction induced by the field as well as the dynamics of trapping in counter-propagating acoustic pulses.
The integral equation formulation leads to equations of motion for the \bubbles\ describing the effect of non-dissipative drag forces.
We show \begin{inparaenum}[(1)]
  \item that the field-induced interactions between the \bubbles\ give rise to effective dipolar interactions, with effective dipoles defined by their velocities, and
  \item that the dominant effect of an ultrasound pulse through a cloud of \bubbles\ gives rise mainly to a translation of the system, though we also observe both expansion and contraction of the cloud determined by the initial system geometry.
\end{inparaenum}
\end{abstract}

\pacs{Valid PACS appear here}

\maketitle

\section{Introduction}

Computational approaches that employ an integral equation formalism to examine acoustic scattering from particles typically assume a static environment in which scatterers remain stationary.
At present, a large body of work details such scattering problems~\cite{Waterman1969, Ding1989, Ye1997}.
While these stationary integral equation methods offer a large degree of accuracy in capturing the underlying physics, many problems of interest require a fully dynamical treatment.
For instance, in biomedical physics, gas-filled \bubbles\ exposed to ultrasonic beams have demonstrated effectiveness as a contrast imaging agent~\cite{Blomley2001} and as drug delivery method~\cite{Allen2002,Hernot2008}, and Ding \etal\ have demonstrated their manipulation using acoustic tweezers in microfluidic channels~\cite{Ding2012}.
Moreover, composite materials consisting of colloidal in-fluid suspensions have peculiar sound propagation properties that can deviate from the ones of homogeneous liquids~\cite{Ye1993}.
In each of these applications, the unconstrained motion of scatterers requires a self-consistent description of their dynamics in conjunction with a description of the acoustic field propagation.

Here, we demonstrate the applicability of coupling particle kinetics to a time-domain integral equation (TDIE) scattering framework to model rigid-sphere motion induced by a time-dependent acoustic potential.
Specifically, we consider the case of an acoustic pulse acting on \bubbles\ that move in a fluid.
Effective Langevin time-averaged radiation pressure forces~\cite{King1934, Borgnis1953}, which consider the case of a steady radiation flux incident upon a body kept in static equilibrium, do not provide an appropriate model in this case as they cannot accommodate inter-particle scattering effects.
While many theoretical and computational descriptions of higher-order acoustic interactions exist~\cite{Gumerov2002, Doinikov2004, Doinikov2005, Ilinskii2007, Azizoglu2009}, few actually make use of computed fields to predict particle trajectories.
As we consider only short-duration pulses, we refrain from time-averaging in favor of using a time-domain scattering formulation to explicitly calculate particle trajectories resulting from a prescribed pulse.
By adopting a weakly-compressible potential formulation of the fluid media, our scalar wave problem inherits a number of similarities and solution techniques from scattering problems in electromagnetic theory, a topic previous works discuss extensively~\cite{Tsang1998,Gumerov2002,Li2014}.
Moreover, our time-domain formulation readily allows the study of transient phenomena (such as acoustic tweezing); a convenience not shared with more common frequency domain approaches.

We structure the remainder of this paper as follows: we first provide a formal mathematical description of the problem---including details on both the kinetic and field methods---followed by data obtained from various pulse and \bubble\ configurations, demonstrating both attractive and repulsive regimes suitable for subtle control of spherical systems in a homogeneous fluid.
Finally, we offer concluding remarks on the effectiveness of the simulation as well as our thoughts on possible future extensions.

\section{Continuum problem statement}

Consider a collection of $N$ rigid, non-intersecting spherical scatterers (\bubbles), each having radius $a_k$, position $\vb{x}_k$, and enclosing volume $V_k \subset \mathbb{R}^3$.
The \bubbles\ move in a homogeneous exterior fluid occupying $V_E$, where we denote the boundary of each \bubble\ as $\Omega_k = \partial V_k$ and thus may ascribe  to each an outward-pointing normal $\vu{n}_k\qty(\theta, \phi)$,
where $\theta$ and $\phi$ represent colatitude and azimuthal angles with respect to the local origin (\bubble\ center).
We wish to investigate the reaction of the system to an incident acoustic pulse, thus the fluid carries a prescribed (band-limited) waveform through the \bubble\ system in which it interacts with each of the $\partial \Omega_k$ according to the ``sound-hard'' regime presented in~\cite{Li2014}. 
The incident acoustic pulse, in combination with the acoustic field scattered from each \bubble\ and the hydrodynamic field induced by the relative velocity of each \bubble, acts as a perturbation to the initially at-rest uniform ideal fluid~\cite{Myers1992, Landau2013}.
We consider here the linear regime, in which the perturbation induced by the acoustic and aerodynamic contribution remain sufficiently small so that the velocity field $\vb{v}(\vb{x}, t)$ satisfies the condition $\abs{\vb{v}(\vb{x}, t)} \ll c_s $, where $c_s$ represents the speed of sound in the fluid.
In this limit, the velocity potential, defined by $\vb{v}(\vb{x}, t) = \nabla \varphi\qty(\vb{x}, t)$, satisfies the scalar wave equation:
\begin{equation}
    \left(\frac{1}{c_s^2} \frac{\partial^2}{\partial t^2}-\nabla^2\right) \varphi\qty(\vb{x}, t) = 0,
  \label{eq:Helmholtz}
\end{equation}
and we may express the pressure perturbation at any point in the exterior medium as
\begin{equation} \label{eq:Bernoulli}
  p\qty(\vb{x}, t) = - \rho_0\pdv{\varphi\qty(\vb{x}, t)}{t},
\end{equation}
where $\rho_0$ denotes the equilibrium density of the fluid.
Rigidity of the $\Omega_k$ necessarily prescribes boundary conditions on the normal velocity components at each interface, namely,
\begin{equation}
  \left.\pdv{\varphi(\vb{x},t)}{\vu{n}_k}\right|_{\vb{x} \in \Omega_k} \mkern-30mu = \dv{\vb{x}_k}{t} \cdot \vu{n}_k.
  \label{eq:boundary_conditions}
\end{equation}
where $\vb{x}_k$ represents the center-of-mass coordinate for the $k^\text{th}$ \bubble.

Using these relations, we apply the Kirchoff-Helmholtz theorem to define the following system of integral equations,
\begin{widetext}
\begin{equation}
  \varphi(\vb{x}, t) = \varphi_\text{inc}(\vb{x}, t) + \sum_{i = 0}^{N - 1}
  \int \dd{t'} \int_{\Omega_k(t')} \mkern-30mu \dd{A}
\qty(
\varphi(\vb{x}',t') \pdv{\GF}{\vu{n}_k} - \GF \pdv{\varphi(\vb{x}',t')}{\vu{n}_k}),
  \label{eq:Kirchoff-Helmholtz}
\end{equation}
\end{widetext}
where $\GF$ denotes the Green's function for a retarded potential, 
\begin{equation}
  \GF = \frac{\delta\qty(t - t' - \abs{\vb{x}-\vb{x}'}/c_s)}{4\pi \abs{\vb{x}-\vb{x}'}}.
  \label{eq:retarded green's function}
\end{equation}
If the system remains localized to a region with small dimensions when compared to the wavelength of sound, retardation effects become negligible and we may instead use the Laplace-kernel Green's function,
\begin{equation}
  G\qty(\vb{x}, \vb{x}') = \frac{1}{4\pi\abs{\vb{x} - \vb{x}'}}.
  \label{eq:green's function}
\end{equation}
To ease notation, we define the following two integral operators,
\begin{subequations}
\begin{align}
  \hat{\mathcal{S}}_k\qty[\varphi(\vb{x} \in \Omega_k(t), t)] &= \int_{\Omega_k(t)} \mkern-30mu \dd{A} G(\vb{x}, \vb{x}') \; \partial_{\vu{n}_k} \varphi(\vb{x}',t) 
  \label{eq:single layer}\\
  \hat{\mathcal{D}}_k\qty[\varphi(\vb{x} \in \Omega_k(t), t)] &= \int_{\Omega_k(t)} \mkern-30mu \dd{A} \varphi(\vb{x}',t) \; \partial_{\vu{n}_k} G(\vb{x}, \vb{x}'), \label{eq:double layer}
\end{align}
  \label{eq:operator_kirchoff}
\end{subequations}
reducing \cref{eq:Kirchoff-Helmholtz} to
\begin{equation}
  \varphi(\vb{x},t) = \varphi_\text{inc} + \sum_{k = 0}^{N - 1} \qty(\hat{\mathcal{D}}_k - \hat{\mathcal{S}}_k)\qty[\varphi(\vb{x} \in \Omega_k(t), t)].
  \label{eq:reduced Kirchoff-Helmholtz}
\end{equation}

In solving \cref{eq:reduced Kirchoff-Helmholtz}, we obtain the velocity potential everywhere for a given time without retarded scattered fields.
For the incident pulse, $\varphi_\text{inc}$, we consider superpositions of wave packets of the form
\begin{equation}
  \varphi_\text{inc}(\vb{x}, t) = P_0\cos(\omega_0 t - \vb{k}\cdot\vb{x})e^{-{(c_s t - \vu{k}\cdot\vb{x})}^2/(2 \sigma^2)}.
  \label{eq:inc_field}
\end{equation}
Finally, the variation in pressure (and thus $\varphi$) over each of the $\Omega_k$ necessarily propels each \bubble\ according to the equation of motion
\begin{equation}
  m_k \dv[2]{\vb{x}_k}{t} = \rho_0 \! \int_{\Omega_k(t)} \mkern-30mu \dd{\vb{S}} \pdv{\varphi\qty(\vb{x}, t)}{t}.
  \label{eq:Newton}
\end{equation}

\begin{figure}[t]
  \centering
  \includegraphics{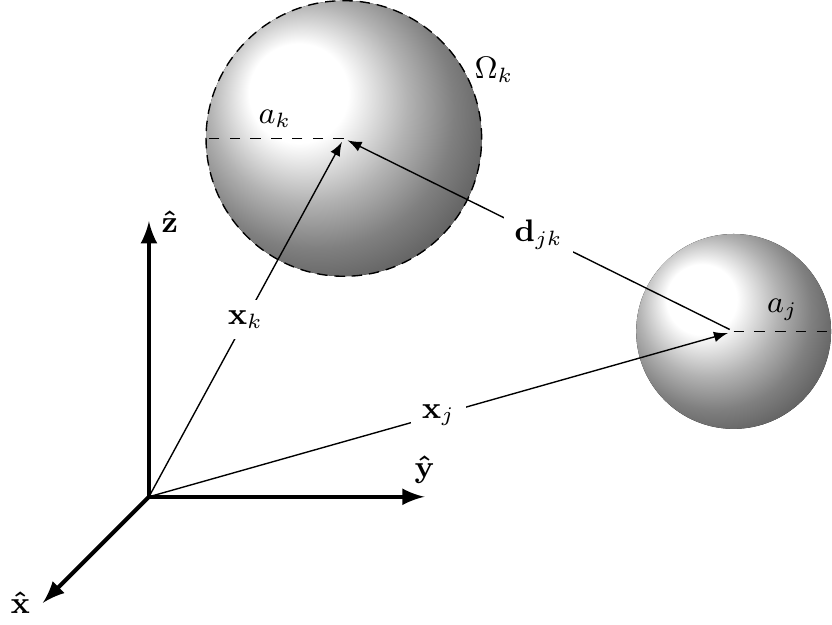}
  \caption{\label{fig:diagram}Coordinate notation.}
\end{figure}

\section{Discretization of the Integral Equations}
To solve the integral equation scattering problem, we begin by discretizing our field in both space and time.
As we have restricted our particles to completely spherical geometries, the spherical harmonics, defined by
\begin{equation} \label{eq:Ylms}
  Y_{\ell m}\quantity(\theta, \phi) = \sqrt{\frac{2\ell + 1}{4\pi}\frac{\quantity(\ell - m)!}{\quantity(\ell + m)!}}P_\ell^m\quantity(\cos\theta)e^{i m \phi},
\end{equation}
give simple eigenfunctions of the operators in \cref{eq:operator_kirchoff}.
As a result, they lend themselves well to an expansion of $\varphi$ on the surface of each \bubble\ with respect to the \bubble's center,
\begin{equation} \label{eq:potential_expansion}
  \varphi(\vb{x}\in\Omega_k, t) = \sum_{\ell \ge 0} \sum_{\abs{m} \le \ell} C^k_{\ell m}(t) \; Y_{\ell m}\qty(\theta, \phi).
\end{equation}
By considering \cref{eq:Bernoulli} and expressing the local velocity potential at each of the $\Omega_k$ as a linear combination of spherical harmonics, we have a complete representation of the body force acting on each \bubble,
\begin{align}
  \vb{F}_\text{body}^k(t) &= - \int_{\Omega_k(t)} \mkern-30mu \dd{\vb{S}} p(\vb{x}\in \Omega_k, t) \nonumber \\
  &= \rho_0 \sqrt{\frac{2\pi}{3}}r^2 \Big(\qty[\dot{C}^k_{11}(t) - \dot{C}^k_{1\, -1}(t)]\vu{x}         \nonumber \\
  &\quad + i\qty[\dot{C}^k_{11}(t) + \dot{C}^k_{1\,-1}(t)]\vu{y} - \sqrt{2}\dot{C}^k_{10}(t)\vu{z}\Big)
\end{align}
due to the orthogonality of dipole terms with the rest of the multipoles.

The problem then becomes one of solving a system of linear equations that we may compactly represent as
\begin{equation}
  \overline{\boldsymbol{\mathcal{Z}}} \cdot \boldsymbol{\varphi} = \boldsymbol{\mathcal{F}},
  \label{eq:z_matrix_system}
\end{equation}
with the overbar denoting a matrix quantity.
We define the elements of $\boldsymbol{\mathcal{F}}$ as projections of the incident field onto local spherical harmonics,
\begin{equation}
  \mathcal{F}_{\ell m}^k = \int_{\Omega_k(t)} \mkern-30mu \dd{A} Y^*_{\ell m}(\theta, \phi) \varphi_\text{inc}(\vb{x}, t),
  \label{eq:forcing projection}
\end{equation}
and detail $\mathcal{Z}^{jk}_{\ell m, \ell' m'}$ for two cases: $j = k$ and $j \not = k$.
In the instances where $j = k$, \cref{eq:operator_kirchoff} propagates effects of the interaction through to every point on a surface sharing a coordinate system with the original, thus the harmonics remain orthogonal and
\begin{equation}
  \mathcal{Z}_{\ell m, \ell' m'}^{jj} = \frac{\ell + 1}{2\ell + 1}  \delta_{\ell \ell'} \delta_{m m'}
  \label{eq:z_matrix_self}
\end{equation}
after exploiting the well-known expansion theorem for \cref{eq:green's function},
\begin{equation}
  G(\vb{x}, \vb{x}') = \sum_{\ell, m}\frac{1}{2\ell + 1}\frac{r_<^\ell}{r_>^{\ell + 1}}Y_{\ell m}\qty(\theta, \phi)Y^*_{\ell, m}\qty(\theta', \phi')
  \label{eq:expansion theorem}
\end{equation}
where $r_< = \mathrm{min}\qty(\abs{\vb{x}},\abs{\vb{x}'})$ and $r_> = \mathrm{max}\qty(\abs{\vb{x}}, \abs{\vb{x}'})$.
A description of the off-diagonal terms where $j \not = k$ proceeds much the same way, though the surface expansions no longer share a local origin, complicating the projections.
Translation operators for the spherical harmonics~\cite{Caola1978,Greengard1987} allow analytic expressions for these matrix elements, though we eschew such operators in favor of numerical integration for speed.

Thus, at every timestep of the simulation, the algorithm proceeds as follows:
\begin{inparaenum}[(i)]
  \item project the incident pulse and surface velocities onto local expansions of spherical harmonics,
  \item propagate scattering effects through space by inverting the operators in \cref{eq:reduced Kirchoff-Helmholtz},
  \item project these scattered fields onto local spherical harmonics to give a total representation of $\varphi$ on each surface, and
  \item move each \bubble\ according to \cref{eq:Newton} \& advance $t \rightarrow t + \Delta t$.
\end{inparaenum}
For rigid \bubbles\ only $\ell = 1$ terms contribute to center-of-mass motion, thus we use only the $C_{1m}$ coefficients in evolving \cref{eq:Newton}.

The inversion in step (ii) above requires some care; by simply inverting the entire propagation operator, $\hat{\mathcal{D}} - \hat{\mathcal{S}}$, to give a single surface pressure, \cref{eq:Newton} reduces to a differential equation of the form
\begin{equation}
  \label{eq:malformed}
  \dot{\vb{x}}_k = f(t, \vb{x}_k, \dot{\vb{x}}_k).
\end{equation}
This presents a number of irregularities with conventional integration schemes and will rapidly diverge towards $\pm \infty$ due to the additional $\dot{\vb{x}}_k$ on the right if implemented \naive ly.
To remedy this, we note that $\hat{\mathcal{S}}$ serves to produce \emph{only} a reaction or drag term on each \bubble\ that impedes motion.
By maintaining quantities for the inversion of $\hat{\mathcal{D}}$ and $\hat{\mathcal{S}}$ separately, we remove the explicit dependence on $\dot{\vb{x}}_k$ by introducing a linear coefficient in the form of an additional mass term---given by the $\dot{\vb{x}}_k$-dependent contribution in the single-layer $\hat{\mathcal{S}}$ operator---when solving \cref{eq:Newton}. 

\section{Analytic results}

\subsection{Single \bubble\ solution}

As an example, consider a single sphere of density $\rho_s$ and radius $a$.
Taking $k a \ll 1$, we may approximate \cref{eq:inc_field} as $\varphi_\text{inc}(\vb{x}, t) = v_0(t) z$ and we wish to find the response velocity of the sphere, $\vb{u}$, in terms of the field velocity $\vb{v} = \nabla \varphi_\text{inc}$. It follows that the expansion of $\varphi_\text{inc}$ contains only $\ell = 1$ terms, thus
\begin{equation}
  \label{eq:uniform field}
  \varphi_\text{inc} = v_0(t) \, a \cos(\theta) 
\end{equation}
on the surface of the sphere. Similarly, from \cref{eq:boundary_conditions},
\begin{align}
  \partial_{\vu{n}} \varphi &= \vb{u} \cdot \vu{n} \nonumber \\
                            &= u_z \, a \cos(\theta)
\end{align}
due to the symmetries present in $x$ and $y$.
As a result, 
\begin{equation}
  \begin{gathered}
    \varphi - \int \dd{S'} \varphi(\vb{x}') \, \partial_{\vu{n}'} G(\vb{x}, \vb{x}') = \\
    v_0 a \cos(\theta) - \int \dd{S}' a u_z \cos(\theta) \, G(\vb{x}, \vb{x}'),
  \end{gathered}
\end{equation}
and it becomes apparent that only $\ell = 1, m = 0$ terms in \cref{eq:expansion theorem} remain after integrating.
Consequently, the field becomes
\begin{equation}
  \label{eq:dipole}
  \varphi(\vb{x}, t) = \qty(v_0(t) \abs{\vb{x}} + \frac{a^3\qty(v_0(t) - u_z)}{2\abs{\vb{x}}^2})\cos(\theta)
\end{equation}
outside the \bubble\ and
\begin{equation}
  \varphi(\vb{x} \in \Omega, t) = \qty(\frac{3}{2}v_0(t) - \frac{1}{2} u_z)a \cos(\theta)
\end{equation}
on its surface.
From this we conclude the total velocity potential in the fluid arises from a surface-scattering term alongside a term describing the transfer of momentum from the moving \bubble\ to the fluid.

Using \cref{eq:Newton}, we may then write the equation of motion for the system as
\begin{equation}
  \rho_s V \dot{u}_z = \rho_0 V \left(\frac{3}{2}\dot{v}_0 - \frac{1}{2}\dot{u}_z\right).
\end{equation}
where $V = 4\pi a^3/3$ gives the volume of the \bubble.
The transfer of momentum from the moving \bubble\ to the fluid becomes a reaction force of the fluid due to the sphere.
Landau \& Lifshitz~\cite{Landau2013} initially derived this non-dissipative \emph{drag force} by way of momentum and energy conservation.
Note that this drag force presents only in the case of accelerated motion of the \bubble\ and we may recast its effect in the form of a \emph{virtual mass} that includes a contribution due to the mass of the displaced fluid,
\begin{equation}
  \qty(\rho_s + \frac{\rho_0}{2})V \dot{u}_z = \frac{3\rho_0 V}{2} \dot{v}_0.
\end{equation}
This expression leads to a simple relation linking $u_z(t)$ and $v_0(t)$ provided the velocity does not remain constant and that the sphere does not move in the absence of the field:
\begin{equation}
  \label{eq:landau result}
  \frac{u_z}{v_0} = \frac{3\rho_0}{\rho_0 + 2\rho_s}.
\end{equation}

The idea of a virtual mass for the accelerated motion of a single sphere in an ideal fluid readily generalizes to the case of a moving collection of mutually-interacting spheres.
Through this, we may compute the dynamics of each \bubble\ in the group, taking into account the effect of the momentum exchange between the fluid and the \bubbles, resulting in both drag and inter-particle forces in addition to the displacement caused by the driving acoustic field.

\subsection{Low-order interactions}

\begin{figure}
  \centering
  \includegraphics{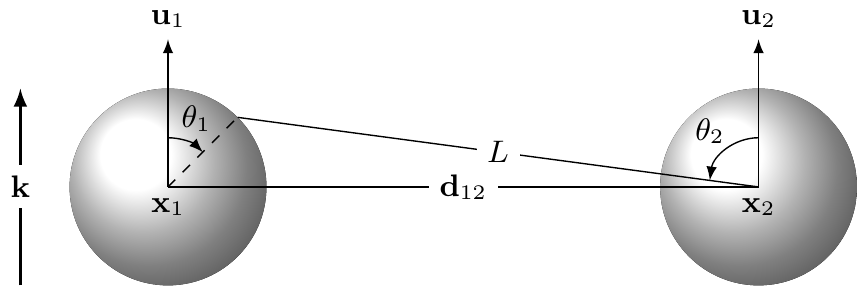}
  \caption{\label{fig:perpendicular}Perpendicular configuration.}
\end{figure}

We now consider two identical \bubbles\ arranged perpendicularly to an incident waveform as in \cref{fig:perpendicular}.
Within the Born approximation, we may take \cref{eq:uniform field} as the incident field and use it in place of the total field on the right-hand side of \cref{eq:reduced Kirchoff-Helmholtz}, assuming negligible contributions from scattering. In doing so, the field everywhere becomes 
\begin{equation}
  \begin{aligned}
  \varphi(\vb{x}, t) = v_0(t) z &+ \frac{a^3}{3} \frac{\cos(\theta_1)}{\abs{\vb{x} - \vb{d}_{12}/2}^2}\qty\big[v_0 - u_1] \\
                                &+ \frac{a^3}{3} \frac{\cos(\theta_2)}{\abs{\vb{x} + \vb{d}_{12}/2}^2}\qty\big[v_0 - u_2].
  \end{aligned}
\end{equation}
By inserting this into \cref{eq:Newton} for $\vb{x}_1$, we have
\begin{equation}
  \label{eq:born newton}
  \begin{gathered}
  m_1 \vb{u}_1 \cdot \vu{z} = 2 \pi \rho_0 a^2 \int \cos^2\theta_1 a^3 \qty(\frac{4}{3}v_0 - \frac{u_1}{3}) \dd{\qty(\cos\theta_1)} + \\
    \rho_0 \int_{\Omega_1} \frac{a^5}{3} \frac{v_0 - u_2}{\abs{\vb{x} - \vb{d}_{12}}^2} \cos\theta_1 \cos\theta_2 \dd{\phi_1} \dd{\qty(\cos \theta_1)}.
  \end{gathered}
\end{equation}
Writing 
\begin{equation}
  \cos\theta_2 = \frac{a}{d_{12}} \frac{\cos\theta_1}{\sqrt{\qty(1 - \frac{a}{d_{12}}\sin\theta_1)^2 + \qty(\frac{a}{d_{12}}\cos\theta_1)^2}}
\end{equation}
and noting $\vb{u}_1 = \vb{u}_2 \equiv \vb{u}_s$ due to symmetry in the initial configuration, we may expand \cref{eq:born newton} in $a/d_{12}$ to give
\begin{equation}
  \label{eq:two bubble born}
  \rho_s u_s = \rho_0\qty(\frac{4}{3}v_0 - \frac{1}{3}u_s) + \frac{\rho_0\qty(v_0 - u_s)}{3}\qty(\frac{a}{d_{12}})^3.
\end{equation}
In the limit of $d_{12} \to \infty$, this becomes
\begin{equation}
  \frac{u_s}{v_0} = \frac{4 \rho_0}{\rho_0 + 3\rho_s}.
\end{equation}
By considering negligible scattered fields at the surface of each \bubble, we qualitatively recover Eq. (27) with different coefficients arising only from the Born approximation. 
Moreover, the additional interaction term in Eq. (31) scales as $\abs{\vb{d}_{ij}}^{-3}$; a behavior anticipated from the dipolar nature of \cref{eq:dipole}.

\section{Numerical Results}

Here we present a series of numerically-solved systems to illustrate the utility of the method in investigating acoustic phenomena.
We perform simulations of one- and two-particle/pulse systems to determine the principal particle-field and particle-particle interactions, followed by simulations of larger assemblages of spheres to investigate group phenomena and effects in systems without symmetry.
Unless otherwise stated, \cref{table:sim parameters} gives the simulation parameters for each of the following simulations; as our interests lie in hydrodynamic applications, we use material parameters characteristic of water to define our external fluid medium.
Similarly, we consider here the case of gas-filled \bubbles~\cite{Blomley2001}, and therefore set their density much smaller than that of the exterior medium. 
The acoustic pulses lie in the ultrasonic regime, and the chosen frequency of \SI{20}{\mega\hertz} corresponds to that of typical applications in acoustic microscopy.

\begin{table}
  \begin{ruledtabular}
    \begin{tabular}{lll}
      Quantity                   & Symbol   & Value                                     \\ \hline
      Sound speed                & $c_s$    & \SI{1500}{\meter \per \second}            \\
      \Bubble\ radius            & $a_k$    & \SI{1}{\micro \meter}                     \\
      Density (exterior)         & $\rho_0$ & \SI{1000}{\kilogram \per \meter \cubed}   \\
      Density (interior)         & $\rho_s$ & \SI{1}{\kilogram \per \meter \cubed}      \\
      Pulse amplitude            & $P_0$    & \SI{0.05}{\meter\squared \per \second}    \\
      Center frequency           & $f_0$    & \SIrange{0.5}{20}{\mega\hertz}            \\
      Pulse duration (st.\ dev.) & $\sigma$ & \SIrange{7}{24}{\micro\second}
    \end{tabular}
  \end{ruledtabular}
  \caption{\label{table:sim parameters}Typical simulation parameters.}
\end{table}

\subsection{Single \bubbles}

\begin{figure}
  \centering
  \includegraphics{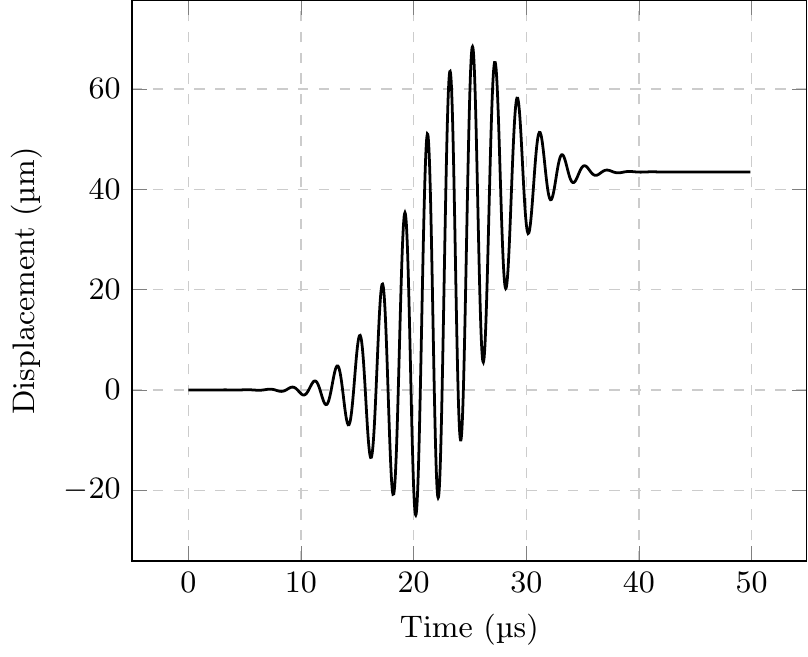}
  \caption{\label{fig:single_displacement}Translation of a single \bubble\ interacting with an incident pulse ($f_0 = \SI{0.5}{\mega\hertz}$, $\sigma = \SI{7}{\micro\second}$). \Bubbles\ interacting with the pulse translate a finite distance along $\vb{k}$ due to the Gaussian envelope in \cref{eq:inc_field}.}
\end{figure}

\Cref{fig:single_displacement} gives the trajectory of a single \bubble\ initially at rest under the effects of an incident Gaussian pulse.
Under the linear and ideal fluid approximations and absent the Gaussian envelope in \cref{eq:inc_field}, the \bubble\ merely oscillates about its origin in accordance with \cref{eq:landau result}.
In the pulsed case, however, the variation in pressure imposed by the finite value of $\vb{k}$ modifies the system dynamics to yield a net translation of each \bubble.
Note that the regime considered here produces no net transfer of momentum between the acoustic field and the \bubble---a consequence of the ideal fluid.

\begin{figure}
  \centering
  \includegraphics{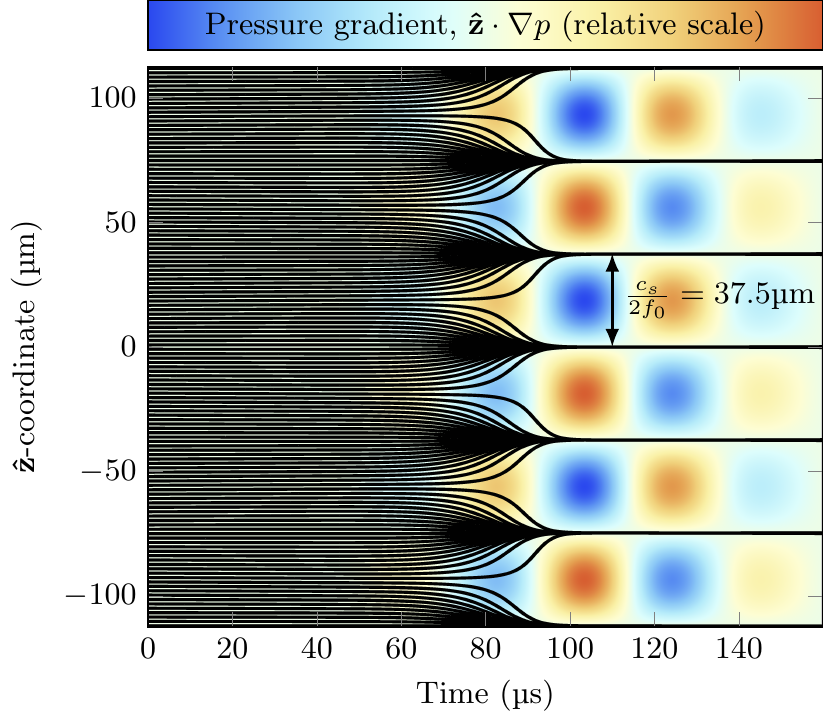}
  \caption{\label{fig:planar_confinement}
  (Color online) Confinement of non-interacting spheres to planes; identical counter-propagating pulses ($f_0 = \SI{20}{\mega\hertz}$, $\sigma = \SI{23.8}{\micro\second}$) initially displaced along $\vu{z}$ tend to align objects in $\nabla P = \vb{0}$ planes at $\lambda/2$ intervals.
    Field \& trajectories sampled every 30 timesteps and smoothed with a 16-sample windowed average.
  }
\end{figure}

\Cref{fig:planar_confinement} depicts smoothed results of 128 trajectories corresponding to single \bubbles\ initially spaced along $\vu{z}$ and excited by identical counter-propagating pulses. 
By taking the width of each pulse much greater than the radius of each \bubble, the two pulses reproduce the effects of interfering standing waves. 
The confinement occurs at $\nabla P = \vb{0}$ (nodal) planes where the net force on each \bubble\ vanishes.
The half-wavelength associated with the dominant pulse frequency gives the separation between neighboring planes.

\begin{figure}
  \centering
  \includegraphics{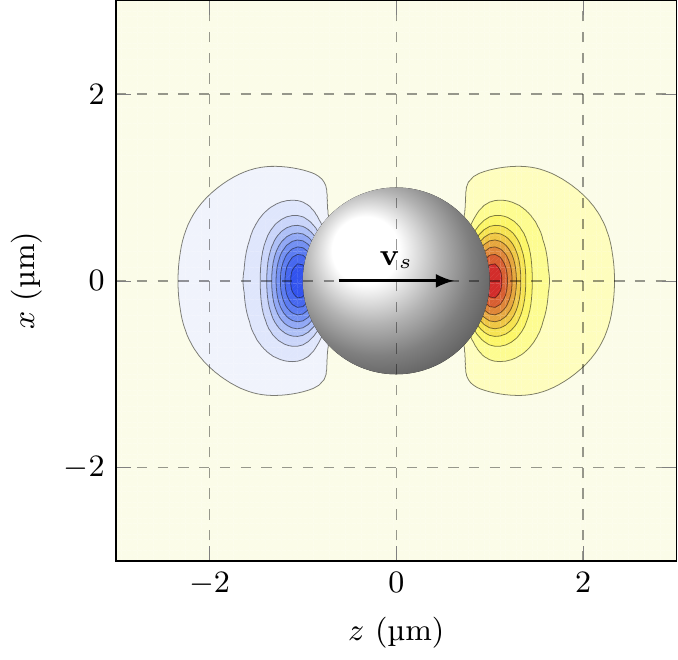}
  \caption{
    \label{fig:dipole field}(Color online)
    Calculated isopotential contours near a lone \bubble.
    Red and blue colorations represent regions of positive and negative potential.
    The motion of each \bubble\ through the background medium serves primarily to produce a dipolar field of velocity potential with $\vb{v}_s$ serving as the sphere's dipole moment.
  }
\end{figure}

Finally, \cref{fig:dipole field} shows the relative velocity potential near a single \bubble; given a surface expansion of $\varphi$, we may compute the potential everywhere through application of \cref{eq:reduced Kirchoff-Helmholtz}.
As predicted by \cref{eq:dipole}, this field greatly resembles that of a
pointlike ``velocity dipole'' with $\vb{v}_s$ acting as a dipole moment.

The simulations described thusfar demonstrate precise acoustic control; through careful application of the incident field parameters, we may induce a (finite, given a finite pulse) translation along the principal $\vu{k}$-vector with a large degree of accuracy in the overall displacement.
In addition, the application of multiple pulses serves to confine \bubbles\ to highly localized regions in space, offering a self-consistent model of acoustic tweezing.

\subsection{Many-particle simulations}

\begin{figure}
  \centering
  \includegraphics{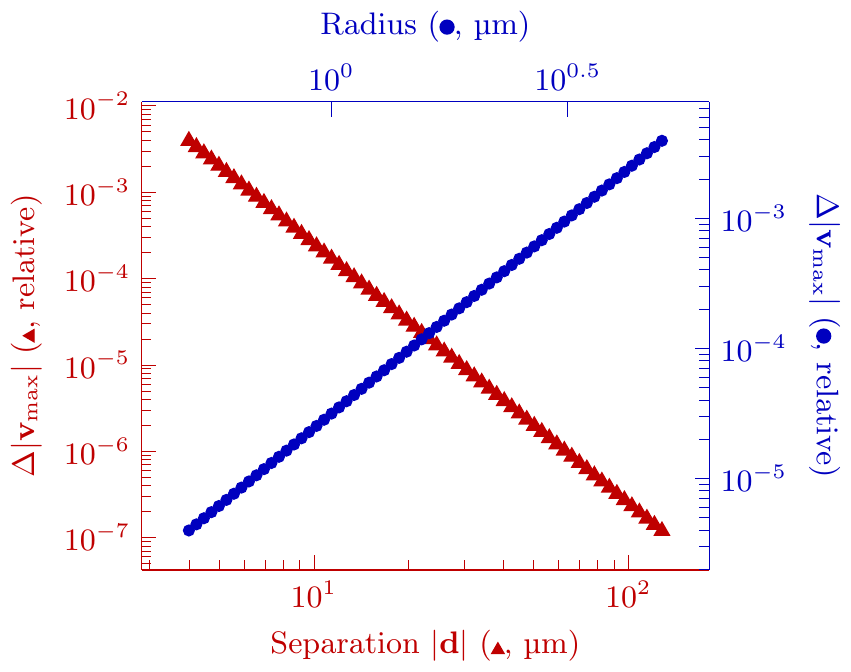}
  \caption{
    \label{fig:double scaling}(Color online)
    Scaling behavior of two \bubbles\ arranged perpendicularly to an incident pulse for various radii and initial separations.
    The (\redTriangle, \blueCircle) symbols on each axis denote data associated with that axis.
    The \redTriangle\ follow a regression of $\Delta \abs{\vb{v}}_\text{d} = \num{0.250754} d_{12}^{-3.00077}$, and the \blueCircle\ follow $\Delta\abs{\vb{v}_\text{max}}_\text{r} = \num{3.13328e-5} a_0^{2.99814}$.
    These trends strongly indicate dominant dipolar interactions between \bubbles.
  }
\end{figure}

\begin{figure}
  \centering
  \includegraphics[width=\columnwidth]{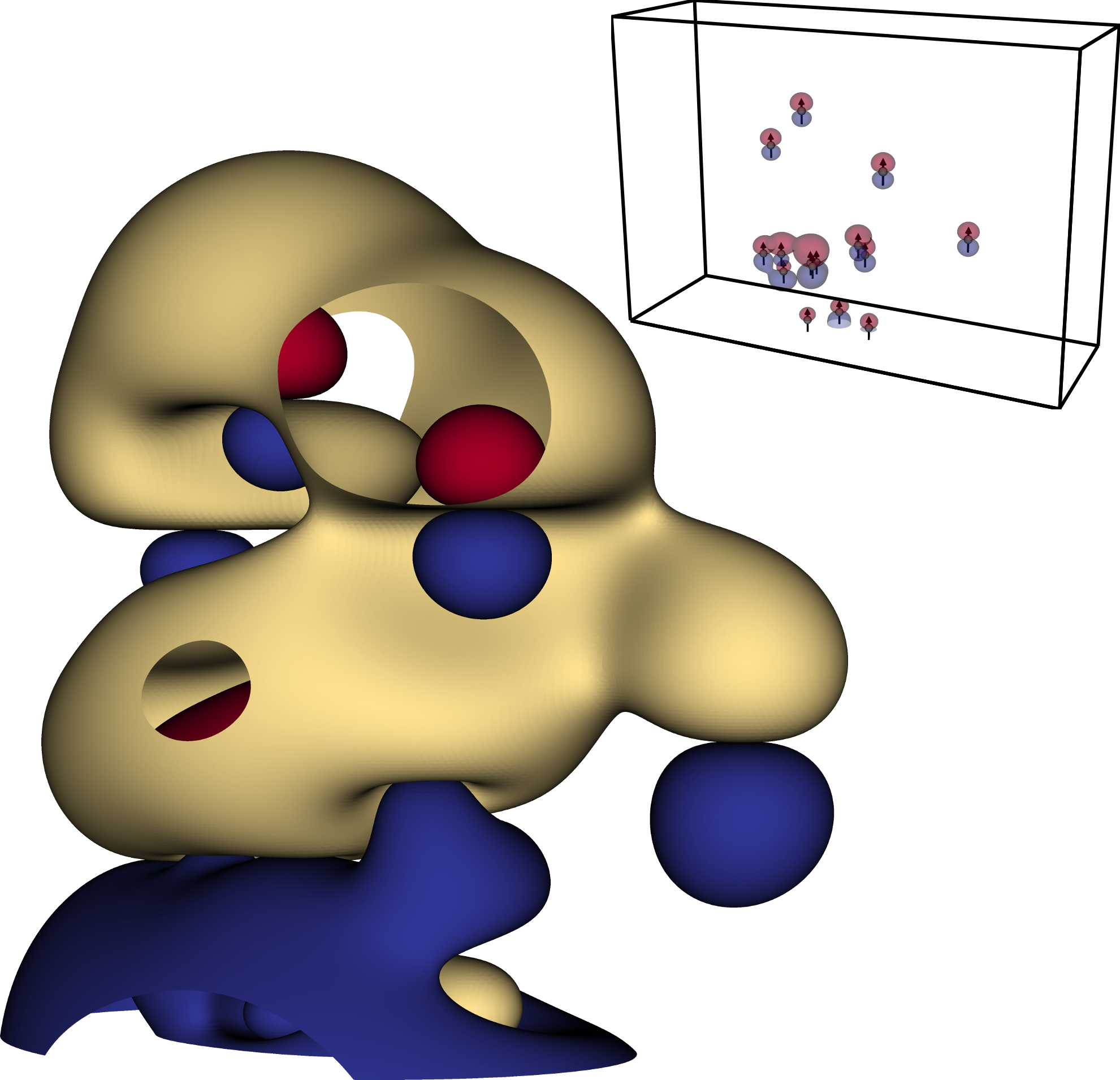}
  \caption{
    \label{fig:isosurface}(Color online) 
    Isosurfaces of velocity potential (arb.~units) calculated by evaluating the $\hat{\mathcal{S}}$ and $\hat{\mathcal{D}}$ terms in \cref{eq:reduced Kirchoff-Helmholtz} for a $N = 16$ particle simulation.
    Red, blue, and yellow surfaces denote regions of positive, negative, and zero potential, with holes appearing due to intersections with the bounding box.
    The inset box shows the three dimensional arrangement of the \bubbles\ superimposed with their velocity vectors, as well as several positive and negative potential isosurfaces.
    Rendered with VisIt\cite{VisIt}.
  }
\end{figure}
We now turn our attention to collections of mutually interacting \bubbles.
To quantify the effects of scattering, we first decouple scattering forces from the incident pulse by arranging two \bubbles\ perpendicularly to the pulse's $\vb{k}$-vector.
\Cref{fig:double scaling} gives results for such a simulation where we plot the relative change in velocity as compared with the single-particle simulation,
\begin{equation}
  \Delta \abs{\vb{v}_\text{max}} = \text{max}\qty(\abs{\vb{v}_\text{double}(t) - \vb{v}_\text{single}(t)}).
\end{equation}
In principle, describing quantities found from a complete simulation as a function of initial separation could obfuscate scaling data considerably; forces arising from scattering could alter the geometry of the system.
In practice, however, the perpendicular configuration used here gives scattering forces that only influence the motion along $\vb{k}$.
Consequently, $\Delta\vb{v} \propto \vb{z}$ and the \bubbles' initial separation remains a good estimator of scaling behavior. 
We see in \cref{fig:double scaling} that the radii data scale as $a_k^3$ and the
separation data exhibit strong $\abs{\vb{d}_{12}}^{-3}$ scaling, again
indicating a dominant dipolar interaction between \bubbles\ as shown by Ilinskii \etal\ in 2007~\cite{Ilinskii2007} and predicted by \cref{eq:dipole}.

\begin{figure}
  \centering
  \includegraphics[width=\columnwidth]{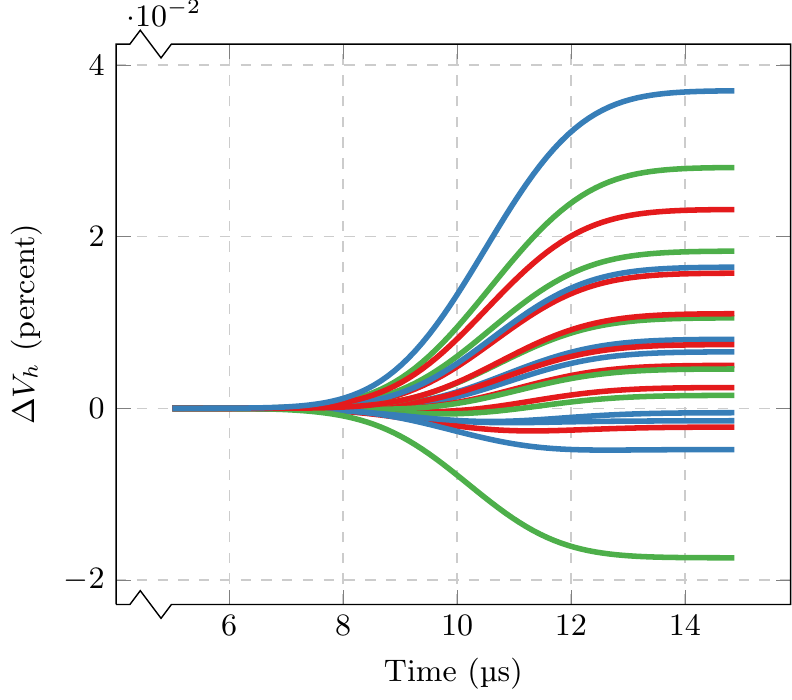}
  \caption{\label{fig:hull change}(Color online)
    Fractional change in the volume of 20 randomly-initialized \bubble\ clouds subject to the same incident pulse, smoothed with a 128-sample moving average.
    Positive and negative values denote expansion and contraction. $\sigma = \SI{1.5}{\centi\meter}$.
  }
\end{figure}

Finally, we consider the dynamics of large ($N=16$) clouds of \bubbles.
For each simulation, we generate a collection of \bubbles\ initialized with zero velocity and random positions within a $\SI{10}{\micro\meter}$ ball subject to a minimum-separation constraint to prevent collisions.
\Cref{fig:isosurface} shows a snapshot of the velocity potential isosurfaces calculated in one such simulation. Even with mutual interactions, the shape of each isosurface remains consistent with the presence of a dipolar field oriented along the \bubbles' velocity.
Again, due to the localization assumption used to justify \cref{eq:green's function}, each system predominantly translates a finite distance in accordance with the results found for a single \bubble\ in \cref{fig:single_displacement}.
To quantify small changes in the geometry of a system, we compute $V_h$, the volume of the convex hull containing each \bubble, at every timestep in the simulation~\cite{SciPy}.
\Cref{fig:hull change} shows the fractional change in the hull volume,
\begin{equation}
  \Delta V_h = \frac{V_h(t) - V_h(0)}{V_h(0)},
\end{equation}
for 20 such systems after smoothing with a weighted moving average.
Curves ending above and below zero indicate larger and smaller hull volumes (system expansion and contraction).
We note from \cref{fig:hull change} a greater tendency for random clouds to expand; the effective dipole-dipole interaction between particles with $\vb{d}_{ij} \perp \vb{k}$ gives purely repulsive forces, while the interaction between particles with $\vb{d}_{ij} \parallel \vb{k}$ gives both repulsive and attractive effects depending on $\sigma$ and the relative phase of the oscillating \bubble\ velocities.

\section{Conclusions}
This work lends a novel, fine-grained approach to the study of acoustic response via integral equation methods.
By considering a potential representation in terms of spherical harmonics on the surfaces of \bubbles\ coupled to a standard molecular dynamics scheme, we obtain a description of the \bubbles' dynamics under the effect of ultrasound pulses without resorting to time-average approximations, though the confined \bubble\ geometries under consideration allow us to neglect small effects arising from time-delays in scattering.
We have shown that the net effect of an ultrasound pulse on a single \bubble\ consists of a translation that we can tune through careful control of pulse parameters.
Additionally, systems with multiple incident waveforms tend to confine \bubbles\ to nodes in the pressure field governed by acoustic interference.
Finally, in the dynamics of systems with many \bubbles, we have observed the effect of weak inter-particle transient effects induced by the driving acoustic pulse.
These effects can produce both expansion and contraction of a cloud of \bubbles, in addition to the overall translation.

Prior work in this area \cite{Zeravcic2011, Tiwari2015} makes use of deformable bubble boundaries about fixed locations. 
Incorporation of these methodologies to our theoretical model naturally offers possibilities for future research, as does the addition of retardation effects. 
Additionally, we expect a straightforward approach to experimental confirmation of the results presented here. 
Optical tracking of tracer particles\cite{Toschi2009} has demonstrated its effectiveness in similar fluid-trajectory studies and would readily adapt to track physical analogues of our theoretical \bubbles.

\begin{acknowledgments}
  The authors would gratefully like to acknowledge NSF grants 1250261 and 1408115 for financially supporting this project as well as iCER and the High Performance Computing facility at Michigan State University for their computational resources.
\end{acknowledgments}

\bibliography{mb2014}

\end{document}